\documentclass[journal,onecolumn,12pt]{IEEEtran} 
\usepackage[]{times} 
\usepackage{epsfig} 
\usepackage{subfigure}
\usepackage{amsmath} 
\usepackage{amssymb} 
\usepackage{graphicx} 
\usepackage{subfigure}
\usepackage{multirow} 
\renewcommand{\baselinestretch}{2.0} 
\newcommand{\bega}{\begin{eqnarray}}
\newcommand{\ega}{\end{eqnarray}}
\newcommand{\bb}{\begin{equation}}
\newcommand{\ee}{\end{equation}}
\newtheorem{defn} {Definition}
\newtheorem{te}{Theorem}
\newtheorem{lema}{Lemma}

\newcommand{\Nat}{I\!\!N}
\begin{document}
\title{Reliable Memories Built from Unreliable Components Based on Expander Graphs}

\author{Shashi~Kiran~Chilappagari,~\IEEEmembership{Student~Member,~IEEE,}
        and~Bane~Vasic,~\IEEEmembership{Member,~IEEE}
\thanks{Manuscript received \today. This work is funded by the NSF under grant CCF-0634969}
\thanks{S. K. Chilappagari is with the Department of Electrical and Computer Engineering, University of Arizona, Tucson, AZ, 85721 USA
(e-mail: shashic@ece.arizona.edu).}
\thanks{B. Vasic is with the Department of Electrical and Computer Engineering, University of Arizona, Tucson, AZ, 85721 USA
(e-mail: vasic@ece.arizona.edu).}}


\maketitle
\thispagestyle{empty}
\begin{abstract}
In this paper, memories built from components subject to transient faults are considered. A fault-tolerant memory architecture based on low-density parity-check codes is proposed and the existence of reliable memories for the adversarial failure model is proved. The proof relies on the expansion property of the underlying Tanner graph of the code. An equivalence between the Taylor-Kuznetsov (TK) scheme and Gallager B algorithm is established and the results are extended to the independent failure model. It is also shown that the proposed memory architecture has lower redundancy compared to the TK scheme. The results are illustrated with specific numerical examples.
\end{abstract}

\section{Introduction}
During the past four decades, the decrease in transistor size and the increase in integration factor have led to very small, fast, and power efficient chips.  As the demand for power efficiency continues, a wide range of new nano-scale technologies is being actively investigated for processing and storage of digital data.  Although it is difficult to discern which of these approaches will become a technological basis for computers in the future, it is widely recognized that due to their miniature size and variations in  technological process, the nano-components will be inherently unreliable.  Even in more traditional semiconductor technologies, reducing transistor size has already started affecting circuit reliability, and it is widely believed that transistor failures (both transient and permanent) will become one of the main technological obstacles as the trend of increasing the integration factor continues. In this paper, we consider storage circuits built from such unreliable (faulty) components. We consider an unreliable component (a logic gate or a memory element) to be a component that is subject to \textit{transient faults}, i.e., faults that manifest themselves at particular time steps but do not necessarily persist for later times  \cite{hadjicostis}. 

Von Neumann \cite{vonneumann} was the first to study computation using faulty gates. In \cite{vonneumann}, he showed that, under certain conditions, increased gate redundancy can  lead to increased reliability of a circuit. However, it was shown that, in general, computation by faulty gates with non-zero computational capacity is not possible (see \cite{dobrushin, pippenger}). The study of storage circuits made of unreliable components led to much more optimistic results. Taylor in \cite{taylor1} proved that a memory has an associated information storage capacity, $C$, such that arbitrarily reliable information storage is possible for all memory redundancies greater than $1/C$. The methodology of the proof, however, does not allow one to explicitly calculate the storage capacity. Taylor considered two models of component failures and proposed construction of fault-tolerant memories based on low-density parity-check (LDPC) codes. In the first model, the failures of a particular component are assumed to be statistically independent from one use to another and is referred to as the independent failure model. In the second model, the components fail permanently but bad components are replaced with good ones at regular intervals. The failures in different components are assumed to be independent in both the models. This construction was further studied by Kuznetsov in \cite{kuznetsov} and we will refer to it as the Taylor-Kuznetsov (TK) scheme. Hadjicostis \cite{hadjicostis} was able to generalize Taylor's scheme to  fault tolerant linear finite state machines. Spielman \cite{spielman2} obtained the best result for a general model of computation, by marrying the ideas of von Neumann with Reed-Solomon (RS) codes.

The fundamental contribution of this paper is to show existence of reliable memories built entirely from unreliable components and which have finite redundancies. We consider the adversarial failure model in which only a fixed fraction of the components fail at any given time and extend our results to the independent failure model using Chernoff bounds \cite{chernoff}. Our memory architecture has lower redundancy compared to the TK scheme. Our fault tolerant memory architecture is also based on LDPC codes but differs from the TK scheme in the decoding algorithm employed. The TK scheme can be shown to be an implementation of the Gallager B decoding algorithm for LDPC codes (the proof will be given in Section \ref{tkscheme}). We use the parallel bit flipping decoding algorithm proposed in the context of expander codes by Sipser and Spielman \cite{spielman}. Expander codes  are a class of asymptotically good error correcting codes with linear time decoding algorithms which can correct a linear fraction of errors. Expander graph based arguments have been successfully applied for message passing algorithms by Burshtein and Miller in \cite{burshtein} as well as for linear programming decoding by Feldman \textit{et.al} in \cite{feldman}.  At the time of their discovery, explicit construction of graphs with expansion required for parallel bit flipping algorithm were not known. Capalbo \textit{et al}. \cite{capalbo} recently gave an explicit construction of expander graphs based on randomness constructors.  Hence, our method can be seen as a constructive proof in contrast to Taylor's method which is an existence proof. 

The rest of the paper is organized as follows. In Section \ref{system} we provide the necessary definitions and a brief overview of LDPC codes. We explain the proposed memory architecture and characterize it in terms of complexity and redundancy. In Section \ref{main} we introduce the model of failure of the components and prove our main result showing the existence of memories which can tolerate failures in all the components. In Section \ref{results} we provide a few numerical examples. In Section \ref{tkscheme} we establish an equivalence between the TK scheme and Gallager B algorithm and extend our results to the independent failure model. In Section \ref{discussion} we discuss open questions and conclude with some interesting remarks.
\section{The System Description}\label{system}
In this section, we give a detailed description of the memory system. We start by introducing the terminology used to characterize memories and proceed to discuss the importance of LDPC codes. We explain the coding scheme and the error correction scheme employed in the proposed memory architecture. We then calculate the redundancy and complexity associated with the memories.
\subsection{Definitions}
A memory is a device in which information is stored at some time and retrieved at a later time \cite{taylor1}. The memories under consideration store information in form of bits and are built from registers (memory elements) each of which can store a single bit. The information storage capability of a memory is the number of information bits it stores. Consider a memory built out of reliable registers. To build a memory with information storage capability of $k$  bits requires $k$ registers. Such a memory is termed as an irredundant memory. Now, consider the problem of information storage with unreliable memory elements. Due to the component failures, the information read out of the memory may not be identical to the information stored originally. Hence, to ensure reliable storage, the information needs to be stored in coded form (see \cite{taylor1} for an excellent discussion on the importance of coded form). Initially, a codeword from some error correcting code is stored in the memory. The unreliable nature of the memory elements introduces errors in the registers and the contents of the memory differ from the initial state. To ensure reliability, a correcting circuit is employed which performs error correction and updates the contents of the registers with an estimate of the original codeword. Hence, a fault-tolerant memory system (referred to as memory system or simply memory henceforth) consists of  memory elements (referred to as storage circuit) and a correcting circuit. The correcting circuit is also built of unreliable components. The coding of information along with the correcting circuit introduce redundancy into the memory system. Such redundant memories are characterized by two closely related parameters, namely, complexity and redundancy. The \textit{complexity} of a memory is the number of components within the memory (a component is a device which either performs an elementary operation or stores a single bit where an elementary operation is any Boolean function of two binary operands \cite{taylor1}). The \textit{redundancy} of a memory is the ratio of the complexity of the memory to the complexity of an irredundant memory which has the same information storage capability \cite{taylor1}. It should be noted that there can be many memory architectures with different complexities but the same information storage capability. 

Another important characteristic of a memory is reliability. We say that arbitrarily reliable information storage is possible in a  memory if the probability of memory failure can be made arbitrarily small. To quantify the reliability of a memory system, it is important to first define what constitutes a memory failure. Let a memory failure be defined as an event in which the word read out of memory is not equal to  the original codeword. Arbitrarily reliable information storage is not possible with such a definition of memory failure. This is due to the fact that the probability of failure is lower bounded by the probability of failure of components in the final step of extracting the information bits. Hence, we define a failure in the following manner. Associated with each codeword in a code is a decoding equivalence class, i.e., the set of words which decode to that particular codeword when decoded with a decoder built of reliable components. If the contents of the memory do not belong to the decoding equivalence class of the original codeword, we say a memory failure has occurred. The \textit{storage capacity}, $C$, of a memory is a number such that for all memory redundancies greater than $1/C$, arbitrarily reliable information storage is possible \cite{taylor1}. 

\subsection{LDPC Codes}
The memories under consideration store information in form of bits and therefore we restrict our attention to binary codes in this paper. An $(n,k)$ binary block code maps a message block of $k$ information bits to a binary $n$-tuple \cite{shulin}. The rate $r$ of the code is given by $r=k/n$. An $(n,k)$ binary linear block code, $\cal{C}$, is a subspace of $GF(2)^n$ of dimension $k$ \cite{shulin}. A parity check matrix $H$ of $\cal{C}$ is a matrix whose columns generate the orthogonal complement of $\cal{C}$, i.e., an element $\mathbf{w}$ of $GF(2)^n$ is a codeword of $\cal{C}$ iff $\mathbf{w}H^T=\mathbf{0}$ \cite{mathworks}. The information storage capability of a memory depends on the type of the code employed in the correcting circuit. Hence, a memory employing an $(n,k)$ block code has information storage capability of $k$ bits.

Taylor in \cite{taylor1} argues that no decoding scheme other than iterative decoding of LDPC codes can achieve non-zero storage capacity. LDPC codes \cite{gallager} are a class of linear block codes which can be defined by sparse bipartite graphs \cite{shokrollahi}. Let $\cal{G}$ be a bipartite graph with two sets of nodes: $n$ variable (bit) nodes and $m$ check (constraint) nodes. The check nodes (variable nodes) connected to a variable node (check node) are referred to as its neighbors. The degree of a node is the number of its neighbors. This graph defines a linear block code of length $n$ and dimension at least $n-m$ in the following way: The $n$ variable nodes are associated to the $n$ coordinates of codewords. A vector $\mathbf{v}=(v_1,v_2,\ldots,v_n)$ is a codeword if and only if for each check node, the sum  of its neighbors is zero. Such a graphical representation of an LDPC code is called the Tanner graph \cite{tanner} of the code. The adjacency matrix of $\cal{G}$ gives $H$, a parity check matrix of $\cal{C}$. An $(n,\gamma,\rho)$ regular LDPC code has a Tanner graph with $n$ variable nodes each of degree $\gamma$ and $n\gamma/\rho$ check nodes each of degree $\rho$. This code has length $n$ and rate  $r \geq 1-\gamma/\rho$ \cite{shokrollahi}. It should be noted that the Tanner graph is not uniquely defined by the code and when we say the Tanner graph of an LDPC code, we only mean one possible graphical representation. 

\subsection{The Proposed Fault Tolerant Memory Architecture}
The complexity and redundancy of a fault-tolerant memory depend on the coding scheme as well as the decoding algorithm employed in updating the contents of the memory. We now explain our memory architecture in detail. 

At time $t=0$, a codeword from an $(n,\gamma,\rho)$ regular binary LDPC code is written into the storage circuit consisting of $n$ registers each of which can store a single bit. The $n$ bits of the codeword correspond to the $n$ variable nodes in the Tanner graph, $\cal{G}$, of the code. The contents of the registers are updated at times $\tau, 2\tau,\ldots, L\tau$, $L \in \Nat$. The update rules can be explained by defining messages along the edges in $\cal{G}$. For a variable node $v$ (check node $c$), let $E(v)$ ($E(c)$) denote the edges incident on $v$ ($c$). Each edge $e$ is associated with a variable node $v$ and a check node $c$. Let $\stackrel{\textstyle \longrightarrow}{\rm {m_{t}}}(e)$ and $\stackrel{\textstyle \longleftarrow}{\rm {m_{t}}}(e)$ represent the messages passed on an edge $e$ from variable node to check node and check node to variable node at time $t$ respectively. Let $v(t)$ denote the value of variable node $v$ at time $t$. Then the update at time $t$ is given by the following algorithm:

\textbf{Algorithm A}
\begin{itemize}
\item For each edge $e$ and corresponding variable node $v$	 
\[
\stackrel{\textstyle \longrightarrow}{\rm {m_{t}}}(e) = v(t^-)
\]
\item For each edge $e$ and corresponding check node $c$
\[
\stackrel{\textstyle \longleftarrow}{\rm {m_{t}}}(e) = \left(\displaystyle\sum_{e'\in E(c) \backslash \{e\}}\stackrel{\textstyle \longrightarrow}{\rm {m_{t}}}(e')\right) ~\mbox{mod} ~2
\]
\item For each variable node $v$

\begin{eqnarray}
v(t^+) &=& \left \{ \begin{array}{cl}
												1,&  \displaystyle\sum_{e\in E(v)}\stackrel{\textstyle \longleftarrow}{\rm {m_t}}(e) > \left\lfloor \gamma/2 \right\rfloor\\
											  0,&  \gamma-\displaystyle\sum_{e\in E(v)}\stackrel{\textstyle \longleftarrow}{\rm {m_{t}}}(e) > \left\lfloor \gamma/2 \right\rfloor\\
											v(t^-), & \mbox{otherwise}  \end{array}\right. \nonumber
											\end{eqnarray}

\end{itemize}	
The algorithm can be interpreted in the following manner. Every variable node sends an estimate of its value to the neighboring check nodes. A check node calculates an estimate of a neighboring variable node by computing the modulo two sum of all the remaining $(\rho-1)$ neighboring variable nodes.
Each variable node receives $\gamma$ estimates, one from each neighboring check and the majority of these estimates is the updated value of the node.

\textit{Remarks:} We assume that the update is instantaneous and use $v(t^-)$ and $v(t^+)$ to denote the value of variable $v$ just before and after the update respectively. We note that the algorithm presented above is a slight modification of the parallel bit flipping algorithm proposed in \cite{spielman}. 

\subsection{Complexity and Redundancy}
LDPC codes can achieve non-zero capacity due to the fact that the redundancy of the LDPC codes memory increases linearly with the information storage capability. The complexity of the logic gates needed to perform decoding depend only on $\gamma$ and $\rho$ and not on the length of the code. So the redundancy remains bounded even as the code length tends to infinity. 

We now calculate the complexity and redundancy associated with our fault-tolerant memory architecture. The storage circuit consists of $n$ registers each of which can store a single bit and hence has complexity $n$. The correcting circuit consists of logic gates (built from components) needed to implement the update algorithm. The message sent from a check node to variable node involves computing the modulo two sum of $(\rho -1)$ bits which requires a $(\rho-1)$-input XOR gate which can be implemented using $(\rho-2)$ two input XOR gates (a two input XOR gate calculates modulo two sum of two bits). Each check node needs to compute $\rho$ such estimates. Therefore, the total number of two input XOR gates is 
\[
(n\gamma/\rho) \times \rho \times (\rho-2)= n \gamma (\rho-2)
\]  
Each variable node is updated based on the majority of the $\gamma$ estimates received from its neighbors. This requires a $\gamma$-input majority logic gate for every variable node whose complexity we denote by $D_{\gamma}$. Hence, the complexity of the memory system is 
\[
\mathcal{S}=n(1+D_{\gamma}+\gamma(\rho-2))
\]
The memory has information storage capability of $rn$ bits and the complexity of an irredundant memory with the same information storage capability is $rn$. The redundancy of the fault-tolerant memory is therefore
\begin{eqnarray*}
R&=&n(1+D_{\gamma}+\gamma(\rho-2))/rn \\ \nonumber
	&\leq&(1+D_{\gamma}+\gamma(\rho-2))/(1-\gamma/\rho) \nonumber
\end{eqnarray*}


\section{Analysis of the Memory System}\label{main}
The storage capacity of a memory depends on the type of failures in the components. A logic gate is said to have failed if its output is flipped. A register is said to have failed if the bit stored in it is flipped. In this paper we consider the adversarial failure model also referred to as bit flipping channel model. In the adversarial model, the failures occur in the worst case fashion but no more than a fixed fraction of the components fail at any given time. In other words, the number of failures is bounded for a given number of components. As the number of components increases so does the number of failures. We denote the fraction of memory element failures in a time interval $\tau$ by $\alpha_{m}$, fraction of two input XOR gate failures for every use by $\alpha_{\oplus}$ and fraction of $\gamma$-input majority logic gate failures for every use by $\alpha_{\gamma}$. As mentioned before, the component failures are transient and independent from one use to another. 

A memory system is said to tolerate a constant fraction of errors in all components if at any time at most a constant fraction of components can fail and no memory failure occurs in the system at all times $t<\infty$. Recall that, from our definition, a memory failure occurs if the contents in the memory do not belong to the decoding class of the originally stored codeword. In this section, we prove that the memory architecture proposed in Section \ref{system} can tolerate a constant fraction of failures in all the components. Our proof is based on the expansion property of the underlying Tanner graph, $\cal{G}$, of the code.  
\begin{defn}\cite{spielman}
A Tanner graph $\cal{G}$ of a $(n,\gamma,\rho)$ LDPC code is a $(\gamma, \rho, \alpha ,\delta)$ expander  if for every subset $S$ of at most an $\alpha n$ variable nodes, at least $\delta|S|$ check nodes are incident to $S$. 
\end{defn}
The definition of expander is much more general but we restrict our attention to Tanner graphs of LDPC codes. 

Sipser and Spielman in \cite{spielman} proposed a class of asymptotically good error correcting codes based on expander graphs known as expander codes. They proposed two simple bit flipping algorithms, namely, serial and parallel and showed that when the underlying graph has sufficient expansion, these algorithms can correct a fixed fraction of errors.  LDPC codes are a special case of expander codes in which the expander graph is the Tanner graph of the LDPC code. 

We describe the parallel bit flipping algorithm and interested readers are referred to \cite{spielman} for details about serial bit flipping. We say that a constraint is satisfied by a setting of variables if the sum of the variables in the constraint is even; otherwise, the constraint is unsatisfied. The set of variable nodes (bits) which differ from their original value are known as corrupt variables.

{\bfseries Parallel Bit Flipping Algorithm}
\begin{itemize}
\item In parallel, flip each variable that is in more unsatisfied than satisfied constraints.
\item Repeat until no such variable remains. 
\end{itemize}

The following theorem from \cite{spielman} gives the sufficient conditions for the parallel bit flipping algorithm to correct a constant fraction of errors.  
\begin{lema}[\cite{spielman}, Theorem 11]\label{thm1}
Let $\cal{G}$ be a $(\gamma, \rho, \alpha, (3/4 +\epsilon)\gamma)$ expander over $n$ variable nodes, for any $\epsilon > 0$. Then, the simple parallel decoding algorithm will correct any $\alpha_0 < \alpha(1 + 4\epsilon)/2$ fraction of error after $\log_{1-4\epsilon}(\alpha_0 n)$ decoding rounds. Also, if $V$ denotes the set of corrupt variables in the input and $|V|<\alpha n (1 + 4\epsilon)/2$, then the parallel decoding algorithm produces a word with at most $|V|(1-4\epsilon)$ corrupt variables after one decoding round.   
\end{lema}
\begin{proof}
See \cite{spielman}
\end{proof}

From Lemma \ref{thm1}, it is clear that a word belongs to the decoding class of a codeword as long as the fraction of corrupt variables (bits) is less than $\alpha(1 + 4\epsilon)/2$. Note that Algorithm A is a slight modification of one iteration of the parallel bit flipping algorithm of \cite{spielman}. In the parallel bit flipping algorithm, every check node indicates to its neighboring variable node if it is satisfied or not. In Algorithm A every check node gives an estimate of the variable node. Theoretically, both the algorithms are equivalent but we use the Algorithm A as it has lesser redundancy. We now state and prove our main theorem.

\begin{te}\label{thm2}
Let $\cal{G}$ be a $(\gamma,\rho,\alpha, (3/4+\epsilon)\gamma)$ expander for any $\epsilon > 0$. The proposed memory architecture can tolerate constant fraction of errors in all the components if 
\[
\alpha_{m}+\gamma(\rho-2)\alpha_{\oplus}+\alpha_{\gamma}<\alpha(1+4\epsilon)(4\epsilon)/2 
\] 
\end{te}
\vspace{3mm}
\begin{proof} At $t=0$, a codeword from an $(n,\gamma,\rho)$ LDPC code with Tanner graph $\cal{G}$ is written into the memory. The contents are updated at times $\tau,2\tau,\ldots,L\tau$, $L \in \Nat$, by running Algorithm A. We bound the number of corrupt variables at time $t$. Let $\alpha_{v}(t)$ denote the fraction of corrupt variables at time $t$. We establish bounds on $\alpha_{v}(t)$ for all $t$. We first prove the following. 
Let $\delta>0$, denote an infinitesimal duration of time. If 
\[
\alpha_{v}((l-1)\tau-\delta)< \alpha(1+4\epsilon)/2,
\] then,
\[
\alpha_{v}(l\tau-\delta) < \alpha(1+4\epsilon)/2 
\]
Let $V(t)$ denote the set of corrupt variables at time $t$. 
\[|V((l-1)\tau-\delta)|=\alpha_{v}((l-1)\tau-\delta)n .\]
Since $\alpha_{v}((l-1)\tau-\delta)< \alpha(1+4\epsilon)/2$, a decoder built with reliable gates outputs a word with at most $|V((l-1)\tau-\delta)|(1-4\epsilon)$ corrupt variables (by Lemma \ref{thm1}). We now bound the number of errors introduced due to the faulty nature of the decoder. Each XOR gate failure can corrupt at most one variable and each majority logic gate failure can corrupt at most one variable. So,
\begin{eqnarray}\label{eq2}
\nonumber |V((l-1)\tau)|&<& |V((l-1)\tau-\delta)|(1-4\epsilon) \\
&+& \gamma(\rho-2)\alpha_{\oplus}n + \alpha_\gamma n
\end{eqnarray}
Eq. \ref{eq2} bounds the number of corrupt variables at the end of $(l-1)^{th}$ correcting cycle. However, in the time interval $[(l-1)\tau~~l\tau)$, at most $\alpha_m n$ variables can get corrupted due to failures in memory elements. Therefore, the time at which there are maximum number of corrupt variables is just before the start of a correcting cycle, i.e.,
\[
\alpha_v(l\tau-\delta)= \mbox{max}\{\alpha_v(t): (l-1)\tau \leq t < l\tau\}
\]
Hence, it suffices to bound $\alpha_v(t)$ for $t=l\tau-\delta, l=1,2,\ldots,L$.
\begin{eqnarray}\label{eq1}
\nonumber |V((l\tau-\delta)|&<& |V((l-1)\tau-\delta)|(1-4\epsilon) \\
&+& \gamma(\rho-2)\alpha_{\oplus}n + \alpha_\gamma n + \alpha_m n
\end{eqnarray}
Dividing Eq. \ref{eq1} by $n$ gives 
\begin{eqnarray*}
\alpha_v((l\tau-\delta)&<& \alpha_v((l-1)\tau-\delta)(1-4\epsilon) \\
&+& \gamma(\rho-2)\alpha_{\oplus}+ \alpha_\gamma  + \alpha_m  \\
&<&\alpha(1+4\epsilon)(1-4\epsilon)/2 + \alpha(1+4\epsilon)(4\epsilon)/2 \\
&=&\alpha(1+4\epsilon)/2\\
\end{eqnarray*}
Since 
\[
\alpha_v(\tau-\delta)\leq \alpha_m < \alpha(1+4\epsilon)/2 ,
\]
it follows that 
\[
\alpha_v(l\tau-\delta)<\alpha(1+4\epsilon)/2 \qquad \forall l \in \Nat .
\]
Hence, 
\[\alpha_v(t) < \alpha(1+4\epsilon)/2 \qquad \forall t<\infty.\]
Since, the fraction of corrupt variables is less than $\alpha(1+4\epsilon)/2$, the contents of storage circuit correspond the decoding class of original codeword and hence, by our definition, no memory failure occurs.
\end{proof}

It is instructive to see the behavior of the memory in the absence of the correcting circuit. In any time interval of $\tau$ seconds, at most $\alpha_{m}n$ fraction of the memories may fail. After sufficiently long time, the fraction of corrupt variables becomes more than $\alpha(1+4\epsilon)/2$ and a memory failure occurs. The presence of a correcting circuit ensures that  at any time the number of corrupt variables remains less than the correcting capability of the code. However, for a given expander there is a loss in the tolerable memory failure due to the faulty nature of the gates as well as the iterative nature of the decoder. Consider the case of where decoder is reliable and failures occur only once. The tolerable fraction of errors for a given expander is close to $\alpha(1+4\epsilon)/2$. In the case of memories with unreliable memory elements but reliable logic gates, the tolerable fraction of memory errors is close to $\alpha(1+4\epsilon)(4\epsilon)/2$. The reduction by a factor of $4\epsilon$ occurs due to the fact that decoder is iterative in nature and needs multiple rounds to converge to the codeword. One round of error correction decreases the errors by a factor of $(1-4\epsilon)$ and $\alpha_m n$ new errors might be introduced due to memory failures. In the extreme case of $\epsilon=1/4$ we have a decoder which takes just one step to correct all the corrupt variables, in which case the tolerable failure rate is arbitrarily close to $\alpha(1+4\epsilon)/2$. The faulty nature of the decoder further reduces the tolerable memory failure rate. Given the values of $\alpha_m, \alpha_{\oplus}, \alpha_{\gamma} $, a code based on graph with sufficient expansion can be chosen to build a fault tolerant memory. It is well known that a random graph is a good expander with high probability (see \cite{spielman} and references therein). In the next section, we illustrate this fact with a few examples.

\section{Numerical Results}\label{results}
In this section, we illustrate with specific numerical examples the redundancies and tolerable failure rates associated with different values of $\gamma$ and $\rho$. We first make the following observations. The redundancy of a memory system depends on the parameters $\gamma$ and $\rho$ of the LDPC code used. Different values of $\gamma$ and $\rho$ can result in same redundancy. To compare across different values of $\gamma$ and $\rho$, the values of $D_\gamma$ and $\alpha_\gamma$ have to be chosen consistently. How $D_\gamma$ and $\alpha_\gamma$ scale with $\gamma$ depends on the technology and implementation. Assuming that all gates are built out of universal NAND gates also does not answer the question fully as different implementations can lead to different values. Hence for the sake of illustration we consider a specific implementation. It should be noted that the subsequent discussion is for illustration purpose only. Accurate analysis for a given case can be carried out along the lines of the method we present in this section. For a given implementation, we fix the the values of $\gamma$ and $\rho$ thereby fixing the redundancy as well as $\alpha_\gamma$ and $\gamma(\rho-2)\alpha_\oplus$. We then use the bounds on the achievable expansion of a $(\gamma,\rho)$ regular bipartite graph to find bounds on the value of $\alpha_{total}=\alpha(1+4\epsilon)(4\epsilon)/2$. This in turn provides bounds on the value of $\alpha_m$ for fixed $\gamma$ and $\rho$. 

\begin{figure*}[htb]
\centering
\subfigure[] 
{
    \label{redundancy9}

\includegraphics[width=0.45\textwidth]{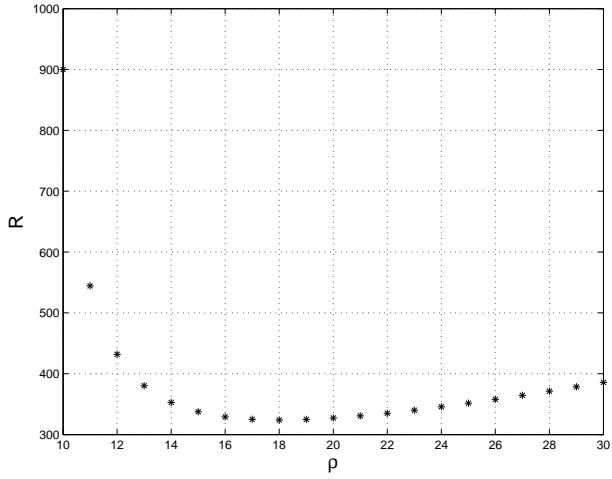}
}
\subfigure[] 
{
    \label{redundancy34}

\includegraphics[width=0.45\textwidth]{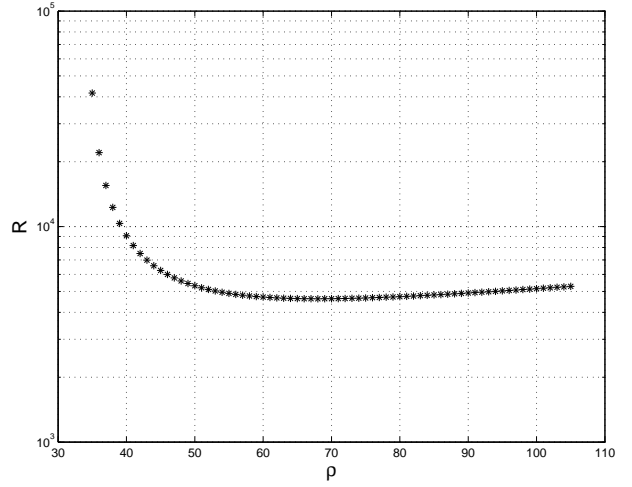}
}
\subfigure[]  
{
    \label{alpha9}

\includegraphics[width=0.45\textwidth]{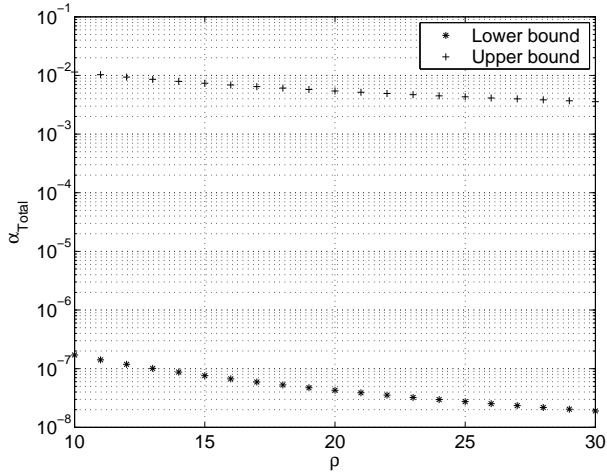}
}
\subfigure[]  
{
    \label{alpha34}

\includegraphics[width=0.45\textwidth]{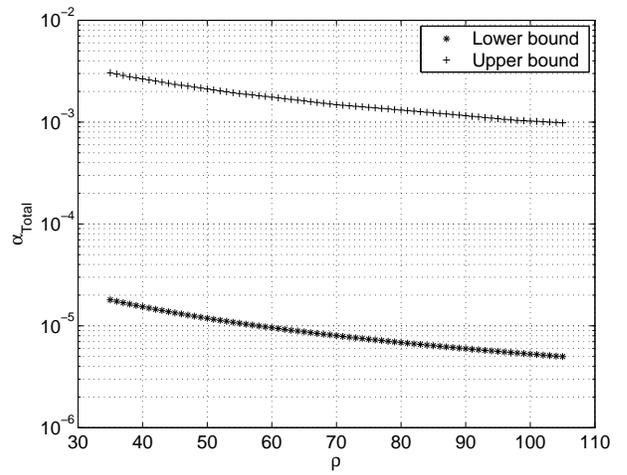}
}
 \caption{Redundancies and bounds on expansion for different values of $\gamma$ \subref{redundancy9} redundancy for $\gamma=9$ \subref{redundancy34} redundancy for $\gamma=34$ \subref{alpha9} bounds on $\alpha_{total}$ for $\gamma=9$ \subref{alpha34} bounds on $\alpha_{total}$ for $\gamma=34$   }
\end{figure*}

\subsection{Redundancy}
Recall that the redundancy of a memory system is given by
\begin{eqnarray*}
R&=&n(1+D_{\gamma}+\gamma(\rho-2))/rn \\ \nonumber
	&\leq&(1+D_{\gamma}+\gamma(\rho-2))/(1-\gamma/\rho) \nonumber
\end{eqnarray*}
For a fixed $\gamma$, $R$ is minimum for a certain $\rho$ depending on the value of $D_\gamma$. For example, if $D_\gamma=2\gamma-1$, then it can be shown that $\rho=2\gamma$ minimizes the redundancy. This implies that a rate $1/2$ code has the least redundancy for a given $\gamma$. 

Fig. \ref{redundancy9} and Fig. \ref{redundancy34} show the dependence of the redundancy on $\rho$ for a given value of $\gamma$.
\subsection{Bounds on Expansion}

We make use of the following theorem from \cite{spielman} to find an upper bound $\alpha_{total}$ for a given $\gamma$ and $\rho$.

[Theorem 25, \cite{spielman}]: Let $B$ be a bipartite graph between $n$ $c$-regular vertices and $(c/d)n$ $d$-regular vertices. For all $0<\alpha<1$, there exists a set of $\alpha n$ $c$-regular vertices with at most
\[
n \frac{c}{d}(1-(1-\alpha)^d) + O(1) \mbox{~neighbors}
\] 
It should be noted that the upper bound is tight for higher values of $c$.

Using this theorem, we can find an upper bound on $\alpha_{total}$ for a given $\gamma$ and $\rho$.  It should also be noted that we look for graphs which expand by at least a factor of $(3/4+\epsilon)$.

The following proposition from \cite{feldman} addresses the issue of existence of expanders.

[Proposition 6, \cite{feldman}]: Let $0 < r < 1$ and $0 < \delta < 1$ be any fixed constants, and let $c$ be such that $(1 - \delta)c$ is an integer which is at least 2. Then for any $n,m$ such that $r = 1 - m/n$ there is a Tanner graph with $n$ variable nodes, $m$ check nodes, and regular left degree $c$ which is an $(\alpha n,\delta c)$-expander, where
\[
\alpha = (2e^{\delta c+1}(\delta c/(1 - r))^{(1-\delta)c)})^{- \frac{1}{(1-\delta)c-1}}
\]
It should be noted that the notation for expanders is different in \cite{feldman}. Also, the proof does not guarantee that all the check nodes have same degree. 

This proposition guarantees the existence of graphs with sufficient expansion and can be used to derive a lower bound on $\alpha_{total}$ for given $\gamma$ and $\rho$. This in turn proves existence of memories which can tolerate $\alpha_m, \alpha_\oplus$ and $\alpha_\gamma$ fraction of failures in respective components as long as $\alpha_m + \gamma(\rho-2)\alpha_\oplus + \alpha_\gamma < \alpha_{total}$. Figs. \ref{alpha9} and \ref{alpha34} illustrate the upper bounds and lower bounds on $\alpha_{total}$ for $\gamma=9$ and $\gamma=34$ respectively. We remark that the bounds have been derived numerically and we do not attempt to give closed form expressions for the bounds as the results are for illustration purpose only. 

\section{The  Taylor-Kuznetsov Scheme and the Independent Failure Model}\label{tkscheme}
As mentioned in the introduction, Taylor \cite{taylor1}, \cite{taylor2} was the first to investigate the capacity and fault-tolerant architectures of storage systems built entirely from unreliable components. His results were refined by Kuznetsov \cite{kuznetsov}. The aim of Taylor and Kuznetsov (TK) was to derive results analogous to the ones derived by Shannon on the capacity of communication systems. The spirit and methodology of Taylor and Kuznetsov's work \cite{taylor1}, \cite{kuznetsov} is similar to Gallager's results \cite{gallager} on LDPC codes.  The bounds on probability of error are given for an ensemble of regular random LDPC codes of infinitely large length used in the correcting circuit. They are obtained under the assumptions that the bits in memory elements remain independent during the process of correction, i.e., under the assumption that the girth (the length of the shortest cycle) of the Tanner graph corresponding to a code is infinitely large. Taylor and Kuznetsov considered a failure model in which a faulty component, generally a logic gate or a memory element, is subject to transient faults, i.e., faults that manifest themselves at particular time steps but do not necessarily persist for later times \cite{hadjicostis}.  It is also assumed that gates fail independently of each other, and that the defects are not permanent, i.e., a gate that malfunctioned at some point in time may give correct output subsequently and that failure occurs by flipping the correct result with some probability $p$, i.e., if the correct result is ``1'', the gate gives ``0'' and vice versa.  Such failure mechanism is referred to as von Neumann type of error or as independent failure model. A faulty gate or memory element in this case can be modeled as as a binary symmetric channel (BSC) with crossover probability $p$.
\subsection{The TK Scheme}
The information to be stored is first encoded by a $(n,\gamma,\rho)$ regular binary LDPC code. The stored codeword $\mathbf{v}=(v_1, v_2,\ldots,v_n)$ consist of bits $v_i, 1 \leq i \leq n$ referred also as variables. Each variable bit $v_i, 1 \leq i \leq n$ is involved in $\gamma$ parity-check equations by $\mathbf{x}H^T=\mathbf{c}$, where $H$ is an $(m \times n)$ parity check matrix and all operations are in binary field. The degree of each check node is $\rho$. The vector $\mathbf{c}=(c_1, c_2,\ldots ,c_m)$ is called syndrome and $c_j$ corresponds to the value of $j^{th}$ parity-check sum for $1 \leq j \leq m$. Parity check $c_j$ is said to be satisfied if $c_j=0$ and unsatisfied if $c_j=1$. A set of parity checks involving bit $x_i$ is $\{c_i^{(1)},c_i^{(2)},\ldots,c_i^{(\gamma)}\}$. After encoding, every coded bit $x_i$ is replaced with $\gamma$ bit-copies of itself $\{x_i^{(1)},x_i^{(2)},\ldots,x_i^{(\gamma)}\}$ and stored in $\gamma$ registers. All bit-copies initially have the same value. New estimates of each of these copies are obtained by using one combination of $\gamma-1$ checks. Note that there are exactly ${\gamma \choose {\gamma-1}}=\gamma$  combinations. The estimates are obtained as follows.
\begin{enumerate}
\item Evaluate parity checks for each bit-copy (exclude one distinct parity check from the original set of checks for each bit-copy).
\item	Flip the value of a particular bit-copy if half or more of the parity checks are unsatisfied.
\item	Iterate (1) and (2).
\end{enumerate}

The Tanner graph description of LDPC codes was unknown at the time of Taylor's paper. It is easy to see that each bit copy corresponds to an edge in the Tanner graph. The variable node corresponding to the edge is the corresponding bit and the check node is the parity check that is excluded in the estimation of that bit copy. If the update scheme is modified so that the check nodes indicate an estimate of the bit copy, then the update rule is an exact implementation of the hard decision message passing algorithm (known as Gallager B algorithm) for iterative decoding of LDPC codes (see \cite{tcaspaper} for a more detailed discussion). Such an equivalence is of great significance as expander graph arguments have been applied to message passing algorithms \cite{burshtein} and allows us to extend these results to the case of unreliable gates also. 

The complexity and redundancy of the original TK scheme are given by
\begin{eqnarray}
\mathcal{S}&=&(2+D_{\gamma-1}+(\gamma-1)(\rho-1))\gamma n \nonumber \\
R &\leq& (2+D_{\gamma-1}+(\gamma-1)(\rho-1))\gamma /(1-\gamma/\rho). \nonumber
\end{eqnarray}

\subsection{The Independent Failure Model}
In this section, we extend our results to the independent failure model. By Chernoff bounds \cite{chernoff}, it follows that a code which can correct a fraction of $p+\Delta$ errors achieves exponentially small probability of error on the BSC with crossover probability $p$ \cite{spielman}. In other words, if there are $n$ components which can fail independently with probability $p$, then the probability that more than $p+\epsilon$ fraction of the components fail at any time is bounded by
\[
\mbox{P(number of failures/n} > p+\Delta) \leq e ^ {-D(p+\Delta || p)n} \leq e^ {-2\Delta^2n}
\]
where $D(x||y)=x \log{(x/y)} + (1-x)\log{((1-x)/(1-y))}$ is the Kullback-Leibler divergence between Bernoulli random variables with parameters $x$ and $y$ respectively.

Now consider a memory architecture built from unreliable components subject to independent failures. Let $p_{m}$ denote the probability of failure of memory element in time interval $\tau$, $p_{\oplus},p_{\gamma}$ denote the probability of failure per use of an XOR gate and a $\gamma$-input majority logic gate respectively. Also, let $\epsilon_{m},\epsilon_{\oplus},\epsilon_{\gamma}>0$ be such that $p_m+\epsilon_m=\alpha_m$, $p_{\oplus}+e_{\oplus}=\alpha_{\oplus}$ and $p_{\gamma}+\epsilon_{\gamma}=\alpha_{\gamma}$. Let $P_f(t)$ denote the probability of memory failure at time $t$. For $\alpha_{m},\alpha_{\oplus},\alpha_{\gamma}$ and $\cal{G}$ satisfying the conditions in Theorem \ref{thm2}, we now have the following theorem 
\begin{te}\label{thm3}
The proposed memory architecture has the following parameters for the independent failure model:
\begin{enumerate}
\item Information storage capability $\geq n(1-\gamma/\rho)$ \label{one}
\item $R \leq (1+D_{\gamma}+\gamma(\rho-2))/(1-\gamma/\rho)$  \label{two}
\item $P_f(L\tau) \leq L e^{(-\Omega(n))}$ \label{three}
\end{enumerate}
\end{te} 
\begin{proof}
(\ref{one}) and (\ref{two}) follow from our discussion in Section \ref{system}. A memory failure may occur if the fraction of components which fail at a time is more than  the tolerable fraction of errors. In $L$ time steps, the correcting circuit is run for $L$ times. The memory registers can fail $L$ times. Hence, we have
\[
P_f(L\tau) \leq L (e^ {-2\epsilon_m^2n} + e^ {-2\epsilon_\oplus^2n} + e^ {-2\epsilon_\gamma^2n}) 
\] 
\end{proof}
The bound on the probability of memory failure given in Theorem \ref{thm3} is a very weak bound and we do not try to improve it. Theorem \ref{thm3} establishes the fact that in the proposed memory architecture, probability of memory failures decreases exponentially with the code length while the redundancy remains bounded. Theorem \ref{thm3} has been stated in the same form as the main theorem in Kuznetsov's paper \cite{kuznetsov}. 

Hence, the proposed memory architecture has exponentially decreasing probability of memory failure in code length and redundancy which is roughly $\gamma$ times less the TK scheme. 

\section{Discussion}\label{discussion}
Taylor in \cite{taylor1} remarks that memories have an associated non-zero storage capacity but an explicit calculation of the capacity is, in general, a difficult problem. For a given failure mechanism, finding storage capacity involves calculating the minimum redundancy to achieve arbitrarily low probability of error. The redundancy is a function of the coding scheme as well as the decoding algorithm. The TK scheme as well as the proposed memory architecture have finite redundancies and  only give bounds on the storage capacity. In this paper, we have shown that there exist reliable memories with redundancies less than that of the TK scheme. This implies that the proposed memory architecture improves the bound by a factor of $\gamma$ at least in a few cases. The explicit calculation of the storage capacity still remains an unsolved problem. While the proposed architecture has less redundancy, the TK scheme may achieve better error exponents as it employs message passing algorithm which is in general more powerful than the parallel bit flipping algorithm. It is worth noting that Taylor in \cite{taylor1} describes the parallel bit flipping algorithm as a scheme for the update rule. He remarks that such an algorithm leads to complex interrelation between the errors as on successive iterations  the values of the bits involved in the estimation of new value of each bit depend on previous value of the bit. We overcome this problem in this paper by using expander arguments. Also, extending the results from the adversarial model to the independent failure model using Chernoff bounds results in very weak bounds on the probability of memory failure.  Using expander arguments directly for the independent failure model for both the proposed architecture and the TK scheme might result in better error exponents as well as lead to tighter bounds on the capacity. 

Another problem which needs to be investigated is the bounds on the probabilities of failures of components, i.e., what are the upper bounds on the  probability of failure of various components. Sipser and Spielman in \cite{spielman} provided explicit construction of codes which can correct a certain fraction of errors. The fraction was later improved by Zemor in \cite{zemor}. Barg and Zemor in \cite{barg} proved that expander codes achieve capacity on the BSC under iterative decoding. Guruswami and Indyk in \cite{guruswami} proposed linear time encodable and decodable codes which achieve optimal error correction performance. Study of fault-tolerant memory architectures based on these codes can provide the required bounds. However, these codes do not directly imply a specific implementation as is the case with parallel bit flipping algorithm. We noted earlier that Capalbo \textit{et al}. \cite{capalbo} gave an explicit construction of expanders. However, the redundancies associated with such expanders are typically very high. This serves as another reason to consider expander codes and other linear time decodable codes based on expanders.

The proposed architecture as well as the TK scheme employ coding scheme based on regular LDPC codes. The works of Richardson, Urbanke and Shokrollahi \cite{richardsonurbankeshokrollahi} and Luby, Mitzenmacher, Shokrollahi, and Spielman \cite{luby} show that well designed irregular codes perform close to capacity. Burshtein and Miller's work on expander graph arguments for message passing \cite{burshtein} is also based on irregular graphs. Investigating memory architectures based on irregular codes may serve as another avenue to study the storage capacity problem.

\section*{Acknowledgment}
The authors would like to thank Milos Ivkovic for fruitful discussions.


\begin{thebibliography}{10}
\providecommand{\url}[1]{#1}
\csname url@rmstyle\endcsname
\providecommand{\newblock}{\relax}
\providecommand{\bibinfo}[2]{#2}
\providecommand\BIBentrySTDinterwordspacing{\spaceskip=0pt\relax}
\providecommand\BIBentryALTinterwordstretchfactor{4}
\providecommand\BIBentryALTinterwordspacing{\spaceskip=\fontdimen2\font plus
\BIBentryALTinterwordstretchfactor\fontdimen3\font minus
  \fontdimen4\font\relax}
\providecommand\BIBforeignlanguage[2]{{%
\expandafter\ifx\csname l@#1\endcsname\relax
\typeout{** WARNING: IEEEtran.bst: No hyphenation pattern has been}%
\typeout{** loaded for the language `#1'. Using the pattern for}%
\typeout{** the default language instead.}%
\else
\language=\csname l@#1\endcsname
\fi
#2}}

\bibitem{hadjicostis}
C.~N. Hadjicostis and G.~C. Verghese, ``Coding approaches to fault tolerance in
  linear dynamic systems,'' \emph{IEEE Trans. Inform. Theory}, vol.~51, no.~1,
  pp. 210--228, Jan. 2005.

\bibitem{vonneumann}
J.~V. Neumann, \emph{Probabilistic Logics and the Synthesis of Reliable
  Organisms from Unreliable Components}, ser. Automata Studies.\hskip 1em plus
  0.5em minus 0.4em\relax Princeton: Princeton University Press, 1956, pp.
  43--98.

\bibitem{dobrushin}
R.~L. Dobrushin and S.~I. Ortyukov, ``Lower bound for the redundancy of
  self-correcting arrangements of unreliable functional elements,''
  \emph{Probl. Inform. Transm.}, vol.~13, pp. 59--65, 1977.

\bibitem{pippenger}
N.~Pippenger, ``Developments in 'the synthesis of reliable organisms from
  unreliable gates','' in \emph{Symposia in Pure Mathematics}, 1990, pp.
  311--324.

\bibitem{taylor1}
M.~Taylor, ``Reliable information storage in memories designed from unreliable
  components,'' \emph{Bell System Technical Journal}, vol.~47, pp. 2299--2337,
  1968.

\bibitem{kuznetsov}
A.~Kuznetsov, ``Information storage in a memory assembled from unreliable
  components,'' \emph{Problems of Information Transmission}, vol.~9, pp.
  254--264, 1973.

\bibitem{spielman2}
D.~Spielman, ``Highly fault-tolerant parallel computation,'' in \emph{IEEE
  Conference on Foundations of Computer Science}, 1996, pp. 154--163.

\bibitem{chernoff}
H.~Chernoff, ``A measure of asymptotic efficiency for tests of a hypothesis
  based on the sum of observations,'' \emph{Annals of Mathematical Statistics},
  vol.~23, pp. 493--507, 1952.

\bibitem{spielman}
M.~Sipser and D.~Spielman, ``Expander codes,'' \emph{IEEE Trans. Inform.
  Theory}, vol.~42, no.~6, pp. 1710--1722, Nov. 1996.

\bibitem{burshtein}
D.~Burshtein and G.~Miller, ``Expander graph arguments for message-passing
  algorithms,'' \emph{IEEE Trans. Inform. Theory}, vol.~47, no.~2, pp.
  782--790, Feb. 2001.

\bibitem{feldman}
J.~Feldman, T.~Malkin, R.~A. Servedio, C.~Stein, and M.~J. Wainwright, ``L{P}
  decoding corrects a constant fraction of errors,'' \emph{IEEE Trans. Inform.
  Theory}, vol.~53, no.~1, pp. 82--89, Jan. 2007.

\bibitem{capalbo}
M.~Capalbo, O.~Reingold, S.~Vadhan, and A.~Wigderson, ``Randomness conductors
  and constant-degree lossless expanders,'' in \emph{STOC '02: Proceedings of
  the thiry-fourth annual ACM symposium on Theory of computing}.\hskip 1em plus
  0.5em minus 0.4em\relax New York, NY, USA: ACM Press, 2002, pp. 659--668.

\bibitem{shulin}
S.~Lin and D.~J. Costello, \emph{Error Control Coding, Second Edition}.\hskip
  1em plus 0.5em minus 0.4em\relax Upper Saddle River, NJ, USA: Prentice-Hall,
  Inc., 2004.

\bibitem{mathworks}
\BIBentryALTinterwordspacing
D.~Terr, ``Parity check matrix.'' [Online]. Available:
  \url{http://mathworld.wolfram.com/ParityCheckMatrix.html}
\BIBentrySTDinterwordspacing

\bibitem{gallager}
R.~G. Gallager, \emph{Low Density Parity Check Codes}.\hskip 1em plus 0.5em
  minus 0.4em\relax Cambridge, MA: M.I.T. Press, 1963.

\bibitem{shokrollahi}
A.~Shokrollahi, ``An introduction to low-density parity-check codes,'' in
  \emph{Theoretical aspects of computer science: advanced lectures}.\hskip 1em
  plus 0.5em minus 0.4em\relax New York, NY, USA: Springer-Verlag New York,
  Inc., 2002, pp. 175--197.

\bibitem{tanner}
R.~M. Tanner, ``A recursive approach to low complexity codes,'' \emph{IEEE
  Trans. Inform. Theory}, vol.~27, pp. 533--547, Sept. 1981.

\bibitem{taylor2}
M.~Taylor, ``Reliable computation in computing systems designed from unreliable
  components,'' \emph{Bell System Technical Journal}, vol.~47, pp. 2339--2266,
  Dec. 1968.

\bibitem{tcaspaper}
B.~Vasic and S.~K. Chilappagari, ``An information theoretical framework for
  analysis and design of nano-scale fault-tolerant memories based on
  low-density parity-check codes,'' \emph{IEEE Trans. Circuits Syst. I, Reg.
  Papers}, accepted for publication.

\bibitem{zemor}
G.~Zemor, ``On expander codes,'' \emph{IEEE Trans. Inform. Theory}, vol.~47,
  no.~2, pp. 835--837, Feb. 2001.

\bibitem{barg}
A.~Barg and G.~Zemor, ``Error exponents of expander codes,'' \emph{IEEE Trans.
  Inform. Theory}, vol.~48, no.~6, pp. 1725--1729, Jun. 2002.

\bibitem{guruswami}
V.~Guruswami and P.~Indyk, ``Linear-time encodable/decodable codes with
  near-optimal rate,'' \emph{IEEE Trans. Inform. Theory}, vol.~51, no.~10, pp.
  3393--3400, Oct. 2005.

\bibitem{richardsonurbankeshokrollahi}
T.~J. Richardson, M.~Shokrollahi, and R.~Urbanke, ``Design of
  capacity-approaching irregular low-density parity-check codes,'' \emph{IEEE
  Trans. Inform. Theory}, vol.~47, no.~2, pp. 638--656, Feb. 2001.

\bibitem{luby}
M.~G. Luby, M.~Mitzenmacher, M.~A. Shokrollahi, and D.~A. Spielman, ``Improved
  low-density parity-check codes using irregular graphs,'' \emph{IEEE Trans.
  Inform. Theory}, vol.~47, no.~2, pp. 585--598, Feb. 2001.

\end{thebibliography}
\end{document}